\documentclass[a4paper]{mem}
\usepackage{natbib}
\usepackage{txfonts}
\usepackage{balance}
\usepackage{graphicx}
\begin{document}

\title{
Production of
Lithium
in Primordial Supernovae
}

\subtitle{}

\author{
  Alexander\ Heger\inst{1-5}
  \and
  Stan Woosley\inst{6}
}

\institute{
  School of Physics and Astronomy --
  Monash University,
  Clayton, Vic 3800, Australia
  \\
  \email{alexander.heger@monash.edu}
  \and
  Tsung-Dao Lee Institute --
  Shanghai 200240, China
  \and
  Center of Excellence for Astrophysics in Three Dimensions (ASTRO-3D), Australia
  \and
  OzGrav-Monash -- Monash Centre for Astrophysics, School of Physics and Astronomy, Monash University, VIC 3800, Australia
  \and
  Joint Institute for Nuclear Astrophysics - Center for the Evolution of the
  Elements (JINA-CEE), Monash University, Vic 3800, Australia
  \and
  Department of Astronomy and Astrophysics --
  University of California at Santa Cruz,
  Santa Cruz, CA 95060, USA
  %\\
  %\email{woosley@ucolick.org}
}

\authorrunning{Heger \& Woosley}

\titlerunning{Lithium from Pop III SNe}

\abstract{

  The first generation of stars is quite unique.  The absence of
  metals likely affects their formation, with current models
  suggesting a much more top-heavy initial mass fraction than what we
  observe today, and some of their other properties, such as rotation
  rates and binarity, are largely unknown or constrained by direct
  observations.  But even non-rotation single stars of a given mass
  will evolve quite differently due to the absence of the metals: the
  stars will mostly remain much more compact until their death, with
  the hydrogen-rich later reaching down ten teems deeper in radius
  then in modern stars.  When they explode as supernovae, the exposure
  to the supernova neutrino flux is much enhanced, allowing for
  copious production of lithium.  This production will not be constant
  for all stars but largely vary across the mass range.  Such
  production even more challenges the presence of the Spite Plateau.

}
\maketitle{}

\section{Introduction}

The tension between the clear predictions for big bang nucleosynthesis
(BBN) production of $^7$Li on the one hand
\citep{2018PhR...754....1P}, and the observation of the much lower
``\textsl{Spite Plateau}'' \citep{1982A&A...115..357S} in metal-poor
or ultra-metal poor (UMP) stars persists to the present day and topic
of many other contributions in these proceedings.  A common
explanation of the past (but see other ideas in these proceedings) has
been to suggest a fixed and universal depletion/destruction process
for $^7$Li from the BBN initial abundance.  In this contribution we
set out to introduce an additional \emph{production} mechanism for
$^7$Li that should be present in the first generation of stars which
also make the metals that are found in the stars of the Spite Plateau.
This adds some extra variation in the initial $^7$Li abundance of the
material of which the Spite Plateau stars have formed.

\section{Massive Population III Stars}

The theory of formation of Population III stars suggest that the first
generation of stars has been quite massive
\citep[e.g.,][]{2002ApJ...564...23B,2002Sci...295...93A,
  2017MNRAS.470..898H}, with typical masses maybe in the tens to
hundreds of solar masses.  Understanding what the initial mass
function (IMF) of the first generation of stars really is, however,
remains an open question.  Here we will assume they are in the massive
star range.

It is this first generation of stars that make the first heavy metals
in the universe, the metals that we find in the most metal-poor
low-mass stars of our galaxy today.  The most iron-poor star found to
date only has an upper limit for its iron abundance
\citep{2014Natur.506..463K,2015ApJ...806L..16B}.  To match the observed
abundances, the best progenitor candidate was a $40\,\mathrm{M}_\odot$
star, with large fallback
\citep{2008ApJ...679..639Z,2018ApJ...852L..19C}, low amount of mixing
\citep{2009ApJ...693.1780J}, and a low dilution factor of $\sim30$ for
the supernova (SN) ejecta with primordial material as the initial
abundance for the ``Keller Star''.  Remember that last number for the
\textsl{Conclusions Sections}.

\begin{figure}[t!]
\includegraphics[width=\columnwidth]{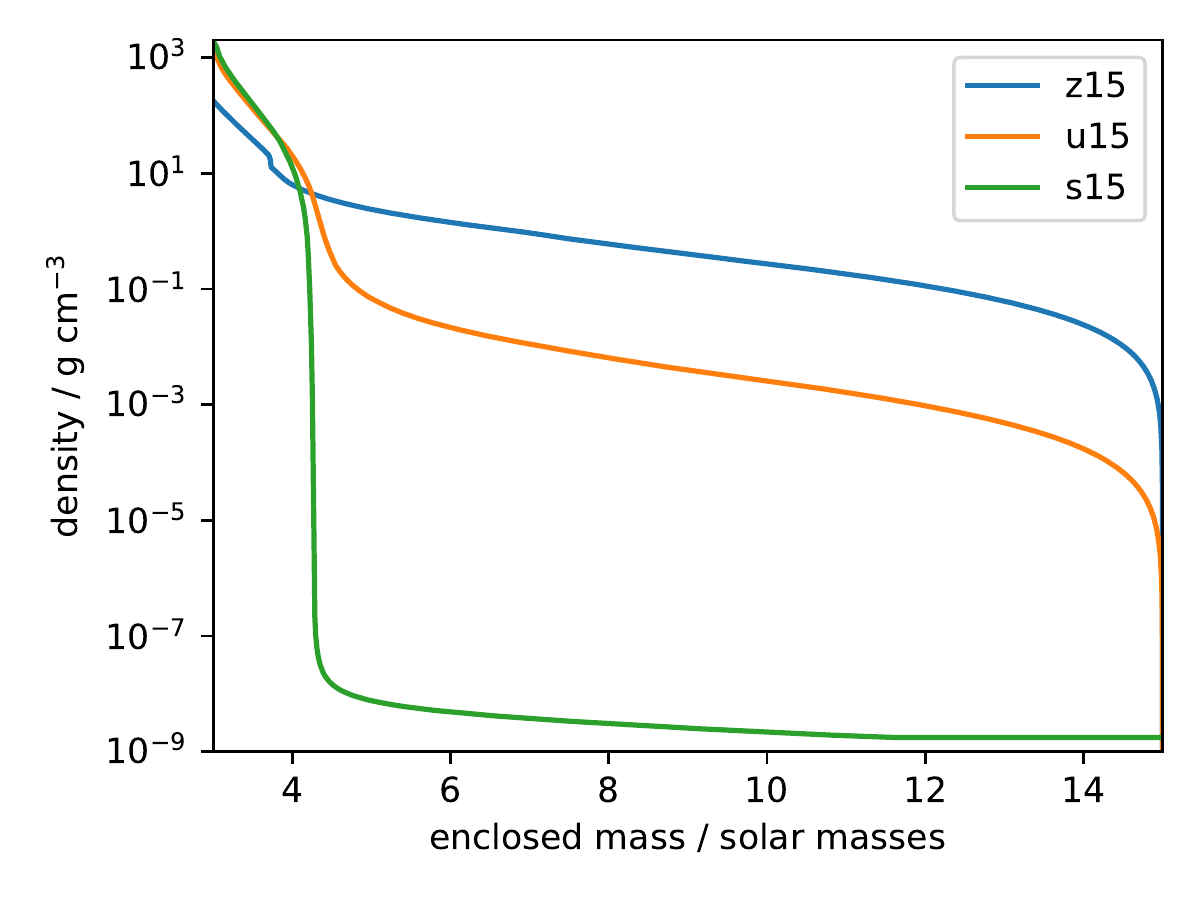}
  \caption{\footnotesize Pre-supernova structure, i.e., density
    ($y$-axis) as a function of mass coordinate ($x$-axis), for
    $15\,\mathrm{M}_\odot$ stars of zero metallicity ($Z=0$,
    \texttt{z15}), $[Z]=-4$ (\texttt{u15}), and solar metallicity
    ($[Z]=0$, \texttt{s15}).  The density in the hydrogen envelope is
    much higher in the Pop III model than in the ultra-metal poor
    (UMP) star or that of solar composition.
\label{dn}}
\end{figure}

\begin{figure}[t!]
\includegraphics[width=\columnwidth]{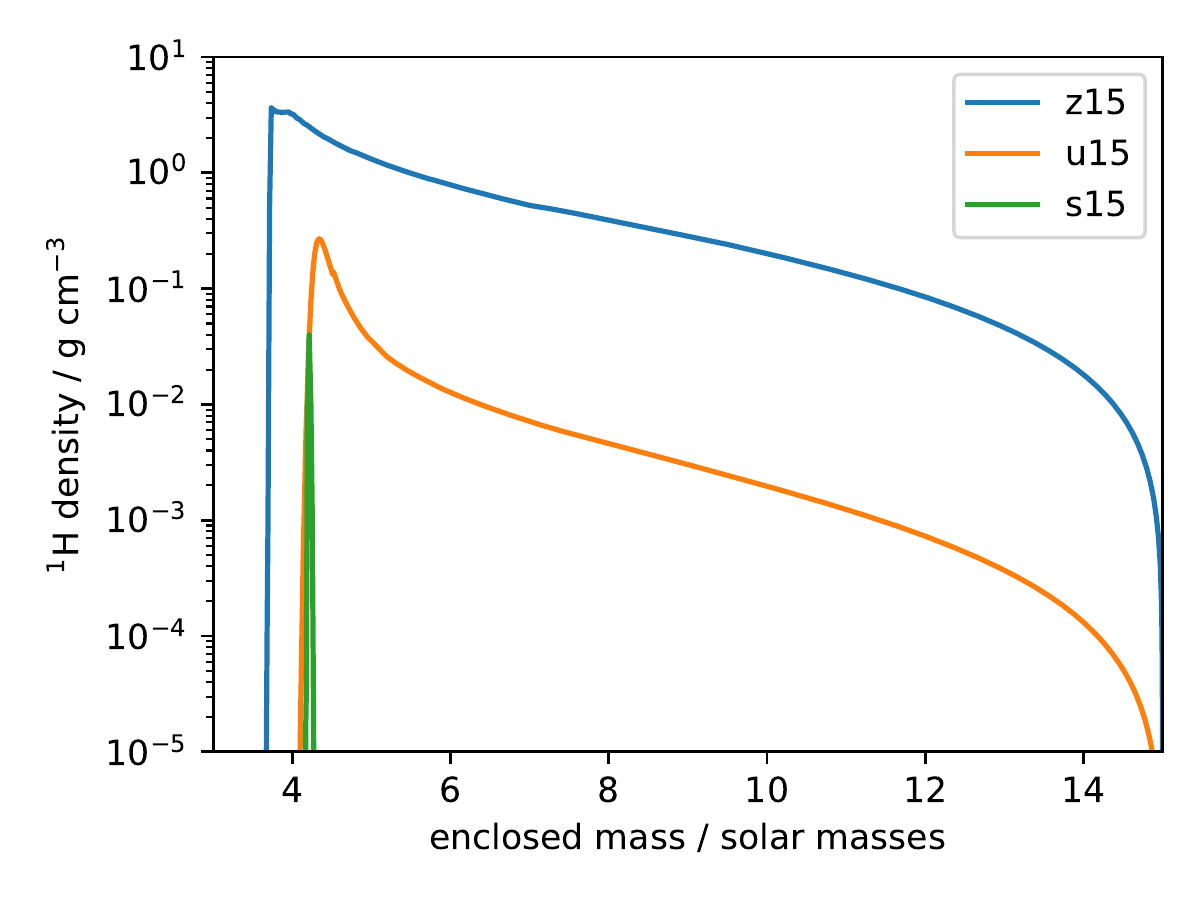}
  \caption{\footnotesize Pre-supernova hydrogen density as a function
    of \emph{mass} coordinate for same models as in Figure~\ref{dn}.
    The hydrogen density in the Pop III star is much higher at the
    base and keeps high density throughout.  For the solar composition
    star only the very bottom mass fraction of the envelope has high
    hydrogen density.
\label{H-M}}
\end{figure}

\begin{figure}[t!]
\includegraphics[width=\columnwidth]{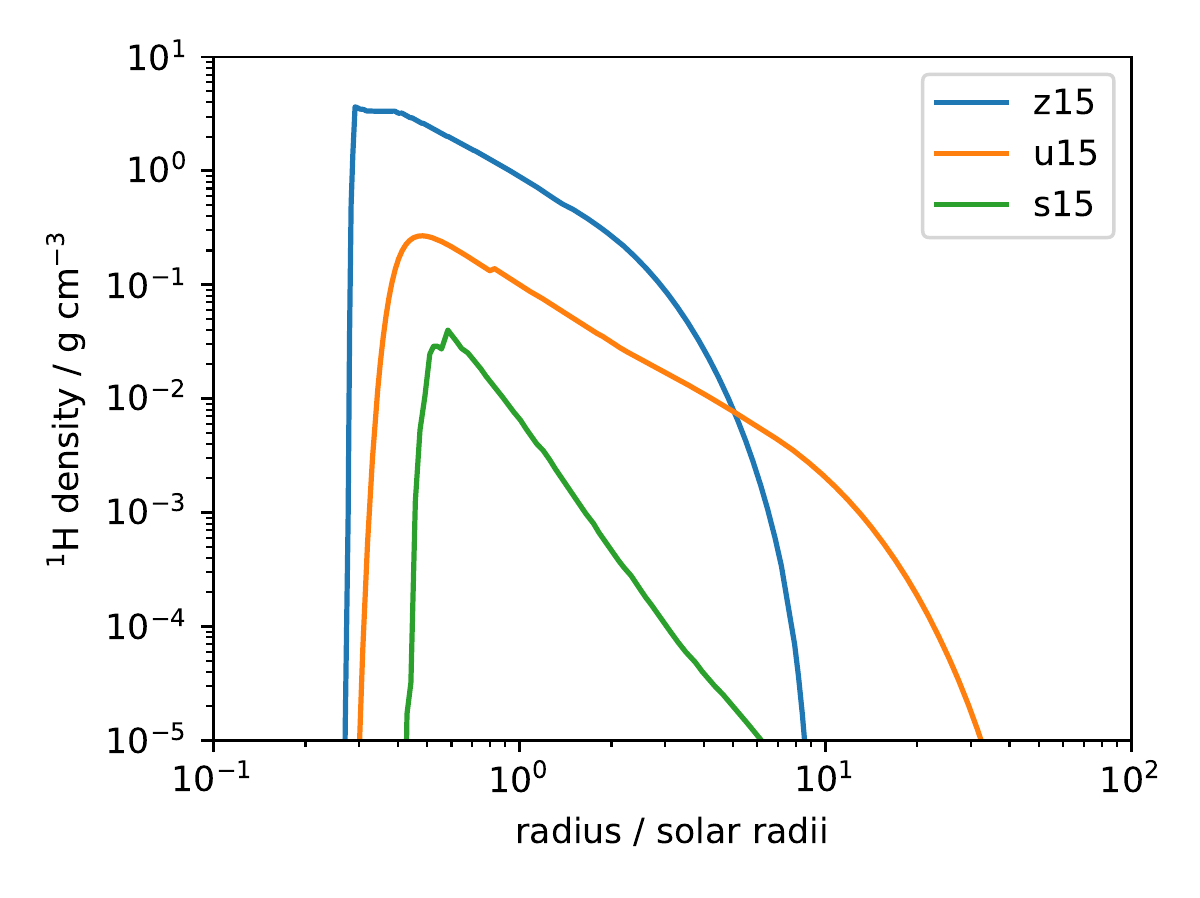}
  \caption{\footnotesize Pre-supernova hydrogen density as a function
    of \emph{radius} coordinate coordinate for same models as in
    Figure~\ref{dn}.  The hydrogen reaches down to lower radii in the
    Pop III star model compared to the UMP star or solar composition.
\label{H-R}}
\end{figure}

\begin{figure}[t!]
\includegraphics[width=\columnwidth]{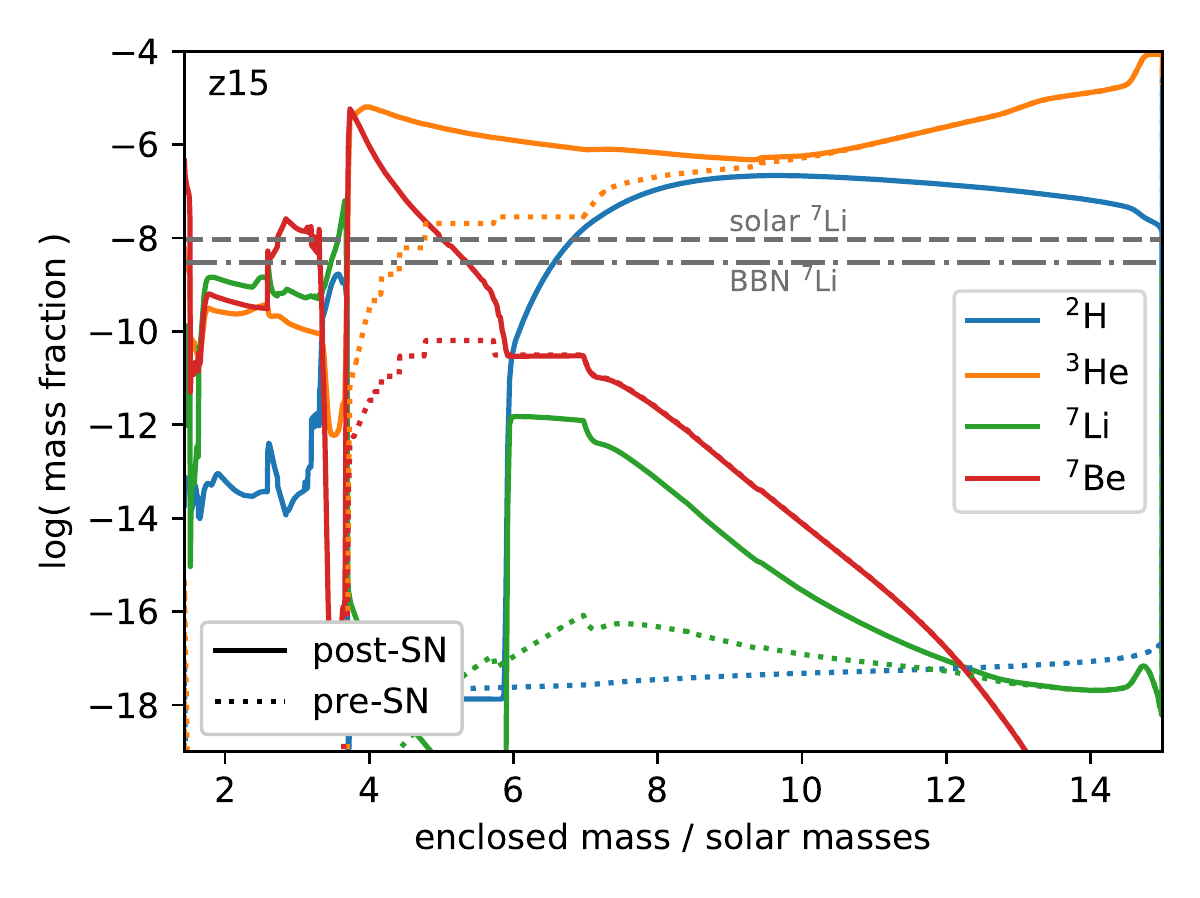}
  \caption{\footnotesize Key abundances at pre-supernova stage
    (\textsl{dotted lines}) and $100\,\mathrm{s}$ after supernova
    explosion (\textsl{solid lines}) of a $15\,\mathrm{M}_\odot$
    primordial star (\texttt{z15}) with an explosion energy of
    $1.2\,\mathrm{B}$.  $^7$Be will still decay to $^7$Li at a later
    time.  We see large production of $^3$He at the base of the
    hydrogen shell (compare to Figure~\ref{H-M}) that burns to $^7$Be
    at the very bottom of the hydrogen shell.  The deuterium that was
    initially made by capture of the neutrons from neutrino
    interaction has all been burnt to $^3$He by the SN shock wave.
\label{Li-z}}
\end{figure}

\begin{figure}[t!]
\includegraphics[width=\columnwidth]{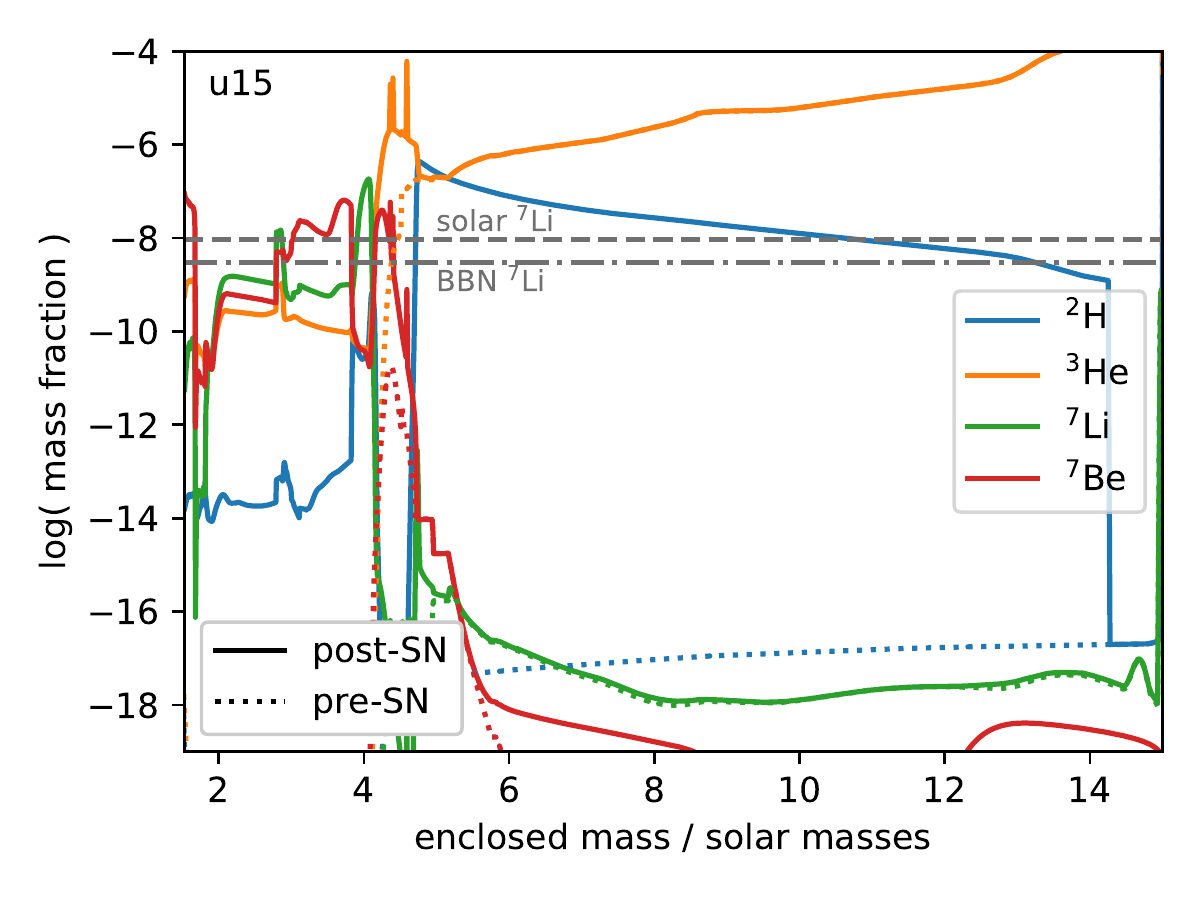}
\caption{\footnotesize Same as Figure~\ref{Li-z} but for Model
  \texttt{u15}. There is production of $^2$H but no significant
  production of $^7$Li or $^7$Be in the hydrogen envelope; there is
  some notable production of $^7$Be from neutrino spallation in the CO
  core.
\label{Li-u}}
\end{figure}

\begin{figure}[t!]
\includegraphics[width=\columnwidth]{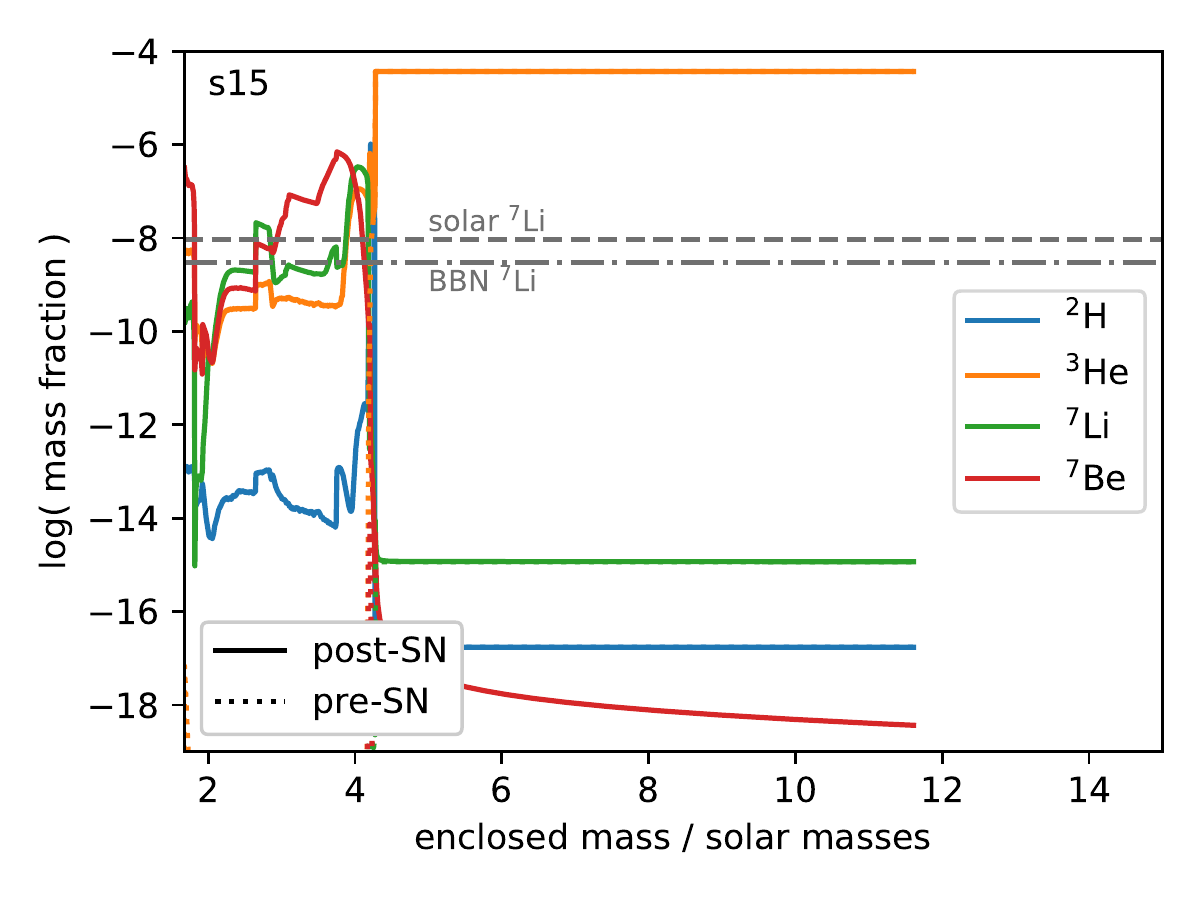}
\caption{\footnotesize Same as Figure~\ref{Li-u} but for Model
  \texttt{s15}.  There is no relevant explosive nucleosynthesis in the
  hydrogen envelope, but some notable production of $^7$Be from
  neutrino spallation in the CO core similar to Model \texttt{u15}.
\label{Li-s}}
\end{figure}

\begin{figure}[t!]
\includegraphics[width=\columnwidth]{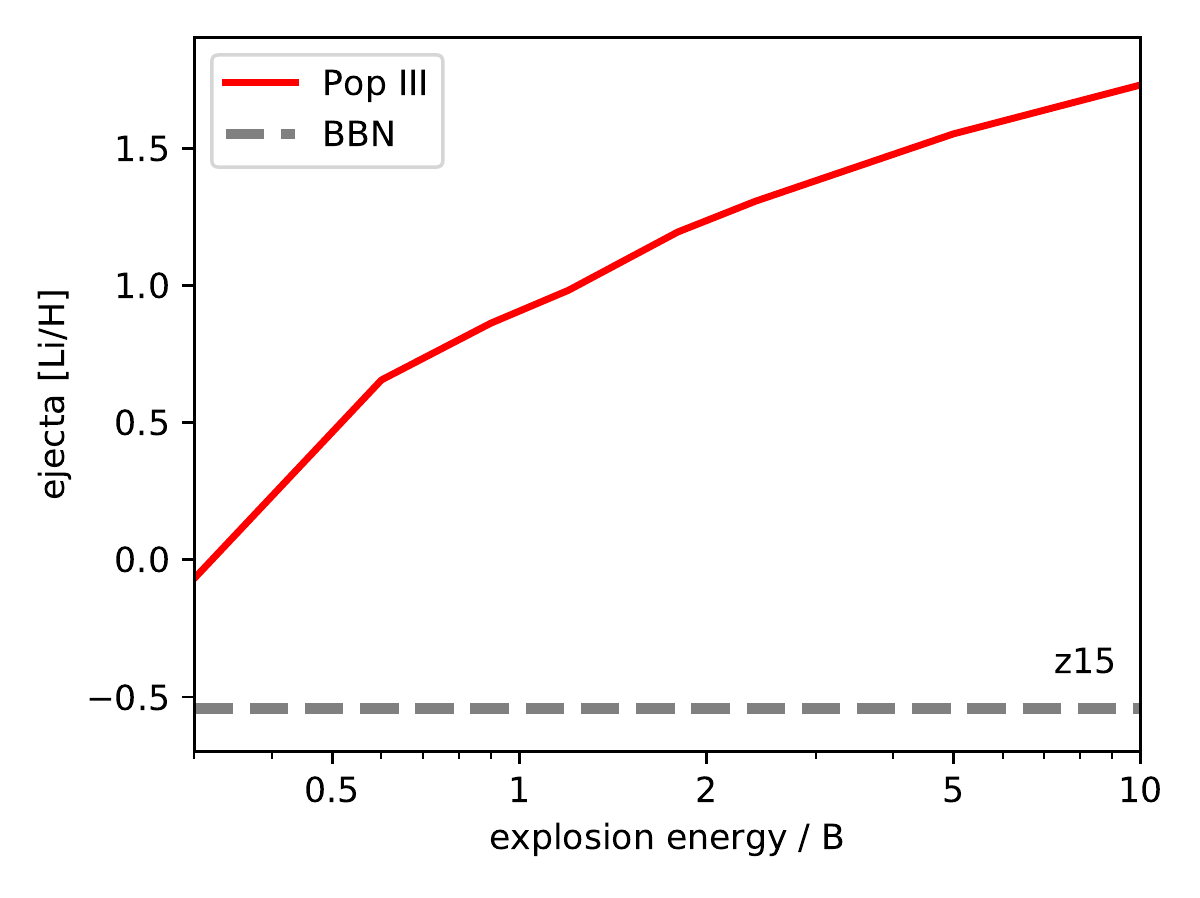}
\caption{\footnotesize Production factor $[\mathrm{Li}/\mathrm{H}]$ in
  the ejecta of a $15\,\mathrm{M}_\odot$ primordial star
  (\texttt{z15}) as a function of explosion energy (final kinetic
  energy of ejecta).  The compression and heating by the explosion in
  the dense hydrogen envelope allows for efficient production of
  $^7$Li by ${}^3\mathrm{He}+{}^4\mathrm{He}\rightarrow{}^7\mathrm{Be}$.
\label{Lie15}}
\end{figure}

\begin{figure}[t!]
\includegraphics[width=\columnwidth]{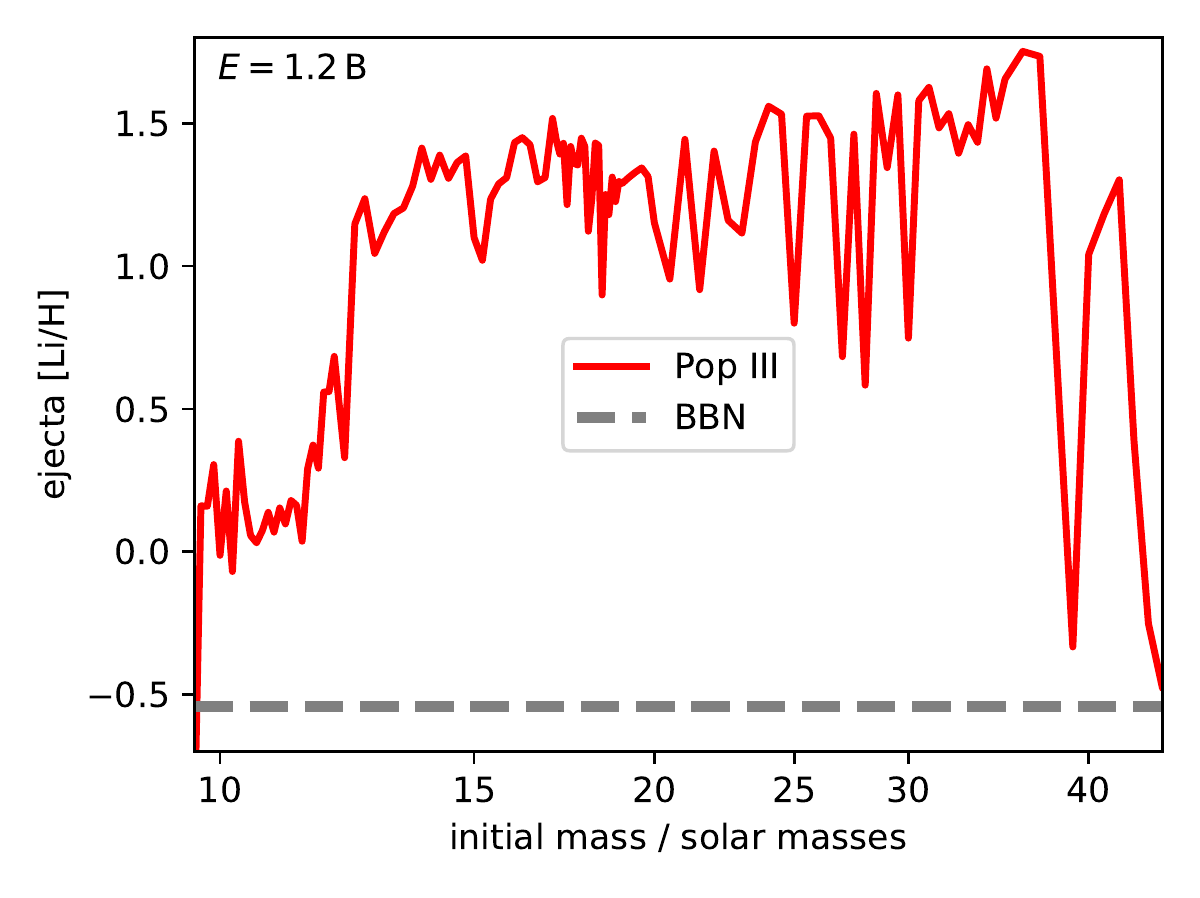}
\caption{\footnotesize Production factor $[\mathrm{Li}/\mathrm{H}]$ in
  the ejecta of a primordial stars with an explosion energy of
  $1.2\,\mathrm{B}$ as a function of initial stellar mass.  For stars
  of higher mass fallback sets in, reducing the $^7$Li yield above
  $40\,\mathrm{M}_\odot$; for higher explosion energies than the
  $1.2\,\mathrm{B}$ case shown here, however, the $^7$Li production
  remains high.
\label{Li-M}}
\end{figure}

\section{Neutrino-Induced Li Production}

Based on the large compilation of non-rotating single Pop III star
models and explosive yields by \cite{2010ApJ...724..341H}, we assessed
the production of $^7$Li in supernovae from primordial stars.  These
stars uniquely lack any initial heavy elements needed for hydrogen
burning by the CNO cycle.  Instead, they contract until the onset of
helium burning by the triple-alpha process, making a small trace of
carbon, just sufficient to start the CNO cycle burning process.  The
same applies to hydrogen shell burning, and, as a result, the entire
hydrogen shell remains quite compact and dense until the supernova
stage as compared to more metal-rich stars (Figure~\ref{dn}).  This
leads to much higher hydrogen density in the envelope
(Figure~\ref{H-M}) and deeper into the star (Figure~\ref{H-R}).

Unique to Pop III stars, the supernova neutrinos can now convert
protons to neutrons that uniquely react onward to make $^7$Li by the
reaction chain
{\small
\[
  {}^1\mathrm{H}\left(\bar\nu_\mathrm{e},\mathrm{e}^+\right)\mathrm{n}
  \left(\mathrm{p},\gamma\right){}^2\mathrm{H}\left(\mathrm{p},\gamma\right)
       {}^3\mathrm{He}\left(\alpha,\gamma\right){}^7\mathrm{Be}\left(\beta^+\right)
       {}^7\mathrm{Li}
       \;,
       \]}
\noindent
leading to high production of $^7$Li in the hydrogen envelope
(Figure~\ref{Li-z}), that is easily ejected.  In contrast, in stars of
higher initial metallicity we do not find $^7$Li production in the
hydrogen envelope, but spallation reactions may produce $^7$Be in the
CO core \citep[shown here in Figures~\ref{Li-u} and
  \ref{Li-s}]{2016EPJWC.10906001B}.  The resulting $^7$Li yield in the
ejecta may still be comparable to the solar abundance level.

Interestingly, the neutrino-induced neutron production alone is not
the only ingredient.  When the supernova shock runs through the dense
hydrogen envelope, the extra heating boost the fusion reaction that
lead to $^7$Li production, and hence there is a sensitivity to
supernova explosion energy, as shown in Figure~\ref{Lie15}.  The
neutrino energy and flux used to compute the neutrino-induced
spallation was kept the same for all explosion energies.

\section{Stellar Mass Summary}

The mechanism presented here operates in supernovae; for star below
the supernova mass limit it does not operate.  We find about solar
production from the onset of supernova explosion up to
$\sim12\,\mathrm{M}_\odot$ (Figure~\ref{Li-M}).  We then find a
plateau of Li production factor of
$[\mathrm{Li}/\mathrm{H}]=1.3\ldots1.5$ from
$\sim12\,\mathrm{M}_\odot$ up to $\sim45\,\mathrm{M}_\odot$ for
explosions of $E=1.2\,\mathrm{B}$.  At higher initial mass, fallback
removes the $^7$Li production, but for higher explosion energies,
e.g., hypernovae, high $^7$Li production would persist provided the
neutrino flux does not cut off, otherwise there would be little
production beyond $\sim45\,\mathrm{M}_\odot$.

We expect no production for very massive stars: For pulsational pair
instability supernovae (PPSN) much of the hydrogen envelope would
either be removed by the pulses \citep{2017ApJ...836..244W} or the
envelope would have much higher entropy and lower density due to the
pulses preceding the final collapse.  For higher masses, the pair
instability supernovae will not make a compact remnant
\citep{2002ApJ...567..532H} and hence not produce a significant
neutrino flux during the explosion \citep{2017PhRvD..96j3008W}.

As a word of further caution, the calculations presented here do not
include rotation or interacting binary star evolution.  Mixing
processes or transfer of enriched material may reduce the $^7$Li
production significantly if it causes a more effective hydrogen
burning by the CNO cycle, which would reduce the density of the
hydrogen envelope.

\section{Conclusions}

Massive Pop III stars can uniquely make $^7$Li by the neutrino-process
due to their exceptionally compact hydrogen envelope that brings the
protons closer in and to higher densities than in later generations of
stars.  The high densities uniquely allow for a reaction chain that
leads to the production of $^7$Be that later decays to $^7$Li.  The
$^7$Li production is induced by charged current reaction on protons
and is therefore sensitive to electron anti-neutrino energies.  The
interaction occurs in the hydrogen envelope outside typical neutrino
oscillation density and can therefore be affected by neutrino flavour
oscillations.  The process is not present in pair instability
supernovae (PSNe) from Pop III stars due to lack of neutrinos nor in
pulsational pair instability supernovae (PPSNe) due to loss of the
hydrogen envelope.

If a dilution factor of $\sim30$, as derived for the Keller Star, is
typical for (some) Pop III SN, the production can be well above big
bang nucleosynthesis for top-heavy initial mass functions (IMF) of Pop
III stars, as is often assumed.  This would result in variations of
the initial $^7$Li abundances of second generations stars such as the
Keller star by a factor 2 or more.  It would make it difficult to explain
the flatness of the Spite plateau by some destruction mechanism acting
on the BBN $^7$Li abundances at an about constant efficiency.  The
origin of $^7$Li abundances in the Spite Plateau would have to be of a
different nature.

Observations of $^7$Li variations in second-generation initial
abundances may provide important constraints on many properties of Pop
III stars, such as internal mixing processes, rotation rates, or their
initial mass function.

\begin{acknowledgements}
  This project has been supported by a grant from Science and
  Technology Commission of Shanghai Municipality (Grants
  No.16DZ2260200) and National Natural Science Foundation of China
  (Grants No.11655002).  Parts of this research were supported by the
  Australian Research Council Centre of Excellence for All Sky
  Astrophysics in 3 Dimensions (ASTRO 3D), through project number
  CE170100013.  Parts of this research were conducted by the
  Australian Research Council Centre of Excellence for Gravitational
  Wave Discovery (OzGrav), through project number CE170100004.  This
  work was supported in part by the National Science Foundation under
  Grant No.\ PHY-1430152 (JINA Center for the Evolution of the
  Elements).
\end{acknowledgements}

\bibliographystyle{aa}
\bibliography{proc}

\end{document}